\documentstyle[12pt]{article}

\begin{document}

\begin{titlepage}

\pagestyle{empty}

\begin{flushright}
{\footnotesize Brown-HET-1011\\
January 1996}
\end{flushright}

\vskip 1.0cm

\begin{center}
{\large \bf Deflationary Models Driven by Matter Creation}
\vskip 1cm
\renewcommand{\thefootnote}{\alph{footnote}}
J. A. S. Lima$^{1,2,}$\footnote{e-mail:limajas@het.brown.edu} and L. R. 
W. Abramo$^{1,}$\footnote{e-mail:abramo@het.brown.edu} 
\end{center}
\vskip 0.5cm

\begin{quote}
{\small $^1$ Physics Department, Brown University, 
Providence, RI 02912,USA.

$^2$ Departamento de F\'{\i}sica Te\'{o}rica e Experimental,
     Universidade \\ 
$^{ }$ $^{ }$ Federal do Rio Grande do Norte, 
     59072 - 970, Natal, RN, Brazil.}
\end{quote}

\vskip 2.5cm

\begin{abstract}

\noindent A nonsingular deflationary 
cosmology driven by adiabatic 
matter creation is proposed. In such 
a scenario 
there is no  
preinflationary stage
as happens in conventional inflationary 
models. Deflation starts from a
de Sitter spacetime characterized by an arbitrary time scale 
$H_{I}^{-1}$, which also pins down 
an initial value for the 
temperature of 
the universe. The model evolves  
continuously towards a slightly modified 
Friedman-Robertson-Walker universe.
The horizon and other well known 
problems of the standard model are 
then solved
but, unlike in microscopic models of inflation, there is no supercooling and 
subsequent reheating. Entropy generation is concomitant 
with deflation and if $H_{I}^{-1}$ is of the 
order of the Planck time, 
the present day value of the radiation 
temperature is deduced. It is 
also shown that the ``age problem" does not 
exist here. In particular, 
the theoretically favored FRW flat model is old 
enough to agree with 
the observations even given the high values 
of $H_{o}$ suggested by 
recent measurements.

\end{abstract}
\end{titlepage}

\pagebreak

\baselineskip 0.6cm

\section{Introduction}

The problem of cosmological particle creation and 
entropy generation is 
presently a very active field of research. It is now 
widely believed 
that matter and radiation need to be created, at least in the very early 
universe, in order to overcome some difficulties presented by hot 
big-bang cosmology [1-17].
  
From a microscopic point of view, such processes were
first investigated by 
Parker and collaborators \cite{Parker} by considering
the Bogoliubov mode-mixing technique in the context of quantum field theory 
in curved space-time\cite{BirrellD}. Despite being rigorous and well-motivated, 
those models were never fully realized  probably, due to 
the lack of a 
well-defined  prescription of how matter creation is to be incorporated 
in the classical Einstein field equations(EFE). This issue has been further explored  in all variants of the inflationary scenario\cite{TraschenRB,Branden}. Usually, the amplification of
zero-point fluctuations of quantum fields by the interaction with the 
background curvature and the corresponding matter creation 
rates are
studied in close analogy with some processes in quantum optics\cite{Branden}.
 
The consequences of matter creation have also been macroscopically 
investigated mainly as a byproduct of bulk viscosity processes near the Planck era as well as in the slow-rollover phase of the new inflationary scenario[5-9].
More recently, the first 
self-consistent macroscopic formulation of the matter creation process 
was  put forward by Prigogine and coworkers \cite{Prigogine} and somewhat 
clarified by Calv\~{a}o {\it et al.} \cite{LCW,CLW}. It was argued that matter
creation, at the expenses of the gravitational field,   
can effectively be discussed in the realm of nonequilibrium thermodynamics. 
In comparison with the standard equilibrium 
equations, the process is 
described by two new ingredients: a balance 
equation for the particle 
number density and a negative  
pressure term in the stress tensor. Such quantities 
are related to each 
other in a very definite way by the second law of 
thermodynamics. In 
particular, the creation pressure depends on the creation rate and may 
operate, at level of the EFE, to prevent either a spacetime 
singularity \cite{Prigogine,AGL} or to generate an early inflationary 
phase \cite{LG,LGA}. 

As it appears,
such an approach may be  
crucial in establishing a pattern for the 
thermodynamics of a viable
quantum-mechanical matter creation process since it seems to incorporate,
in a rather simple scheme, the backreaction contributions.
In this formulation, the laws of nonequilibrium thermodynamics  
were used since the very beginning, thereby leading to definite relations 
among classical quantities and, more important, establishing  the 
temperature evolution law for the created particles. In this 
way, the fragile semiclassical analysis needed by 
the former models of 
matter creation  as well as by some analog 
treatments, such as the one 
represented by bulk viscosity mechanisms, lose 
significance \cite{LG,Gabriel}. The 
stress tensor for matter creation now has a 
well-defined expression which
depends on the usual physically meaningful quantities.

Inflation, on the other hand, has also been invoked as a key aspect of 
any viable cosmological
scenario, since in its latest versions it could explain several
puzzles of the visible universe: the horizon problem,  the entropy 
of the cosmic microwave background (CMBR) and 
the causal generation of seeds to 
form the large-scale 
structure\cite{BR 95}. It is interesting to 
note, however, that the inflationary 
scenario is not free of their own drawbacks, which 
seem to be a strong 
evidence  that such a mechanism is probably only a piece of a more 
complete cosmological theory. For instance, in some
inflationary variants, there is a preinflationary 
stage in which the universe 
emerges from a radiation-dominated FRW phase and, due to the vacuum 
dominance, enters in the de Sitter epoch
at a given critical temperature. This means that inflation does not 
evade the singularity problem, where matter and 
radiation  were {\it ab initio} created in a 
single event at $t=0$. In fact,
as shown by Borde and Vilenkin\cite{BV 94}, if some 
reasonable physical 
conditions are satisfied such models must necessarily possess an initial singularity. On the other hand, 
due to adiabatic expansion of the de Sitter
phase, the universe undergoes a supercooling 
in which the temperature 
decreases by a factor of $10^{-28}$ in order to solve the horizon 
problem. Subsequently, 
it is reheated through a highly nonadiabatic 
process during the coherent 
oscillations of the inflaton field\cite{TraschenRB}. In this connection, it is worth mentioning 
that there is neither a coherent kinetic 
theoretical treatment nor a 
nonequilibrium thermodynamics formulation  
describing  phenomena very 
far from equilibrium. Another well-known difficult of inflation is 
related to the ``age problem": the theoretically favored 
FRW flat model has an expansion 
age of $\frac {2}{3}H_{o}^{-1}$ which by
the latest measurements \cite{Pierce,Freedman} gives only 8.3 Gyr or less. 

In this context, the present article discusses a class of deflationary cosmologies endowed with ``adiabatic'' matter creation. By deflation we mean inflation without a preinflationary stage i.e. the
beginning of the universe in the remote past is
like a pure de Sitter spacetime, irrespective of the curvature parameter. The ``adiabatic'' condition characterize thermodynamically the creation process.
In this way, particles(and consequently entropy) are continuously generated, however, the specific entropy per 
particle remains constant during the
process\cite{LCW,CLW}. In the case of photons this also means that the equilibrium relations are preserved($n \sim T^3, \rho \sim  T^4$) and that the 
photon spectrum is compatible with the present
isotropy of the cosmic background radiation\cite{JA 96,JA1 96}. 

As we shall see, due to 
the existence of a long period of deflation, the aforementioned problems 
are solved in a natural way. Unlike inflation, this model
predicts that the de Sitter phase slowly cools down, smoothly 
connecting to a FRW-like evolution scheme.
One clear advantage of such a model is that the universe is
always in quasi-equilibrium, as opposed to having 
several ``hiccups'' of
extremely out-of-equilibrium phase transitions. In 
particular, the rapid 
supercooling followed by the dramatic reheating existing in the 
microscopic models have been avoided. The model is nonsingular with
the cosmic history starting from a de Sitter state 
characterized by
an arbitrary time scale $H_{I}^{-1}$. If $H_{I}^{-1}$ is of the
order of the Planck time, the present day 
temperature of the CMBR can be 
derived. It is also shown that the age of 
the universe is large enough to 
agree with observations even given the high value of $H_{o}$ suggested by 
the latest measurements.

\section{Basic Equations}

We start with the homogeneous and isotropic FRW line element
\begin{equation}
\label{line_elem}
  ds^2 = dt^2 - R^{2}(t) (\frac{dr^2}{1-k r^2} + r^2 d\theta^2+
      r^2sin^{2}(\theta) d \phi^2) \quad ,
\end{equation}
where $R$ is the scale factor and $k= 0, \pm 1$ is the curvature 
parameter. Throughout we use units such that $c=1$.

    In that background, the nontrivial EFE for a fluid endowed with 
matter creation and the balance equation for the particle number 
density can be written as\cite{CLW,LG,LGA}

\begin{equation}
    8\pi G \rho = 3 \frac{\dot{R}^2}{R^2} + 3 \frac{k}{R^2} \quad ,
\end{equation}

\begin{equation}
   8\pi G (p+p_{c}) = -2 \frac{\ddot{R}}{R} - \frac{\dot{R}^2}{R^2} -
	\frac{k}{R^2} \quad ,
\end{equation}

\begin{equation}
      \frac{\dot{n}}{n} + 3 \frac{\dot{R}}{R} = 
           \frac{\psi}{n}
               \quad ,
\end{equation}
where an overdot means time derivative and $\rho$, $p$, $p_{c}$, $n$ and
$\psi$ are the energy density, thermostatic pressure, creation presure,
particle number 
density and matter creation rate, respectively. The creation pressure 
$p_{c}$ is defined in terms of the creation rate and other 
thermodynamic variables. In the case of adiabatic 
matter creation, it is given by
 
\begin{equation}
    p_{c} = - \frac{\rho + p}{3nH} \psi ,
\end{equation}
where $H = {\dot {R}}/R$ is the Hubble parameter.

By combining the usual ``gamma-law'' equation of state 

\begin{equation}
\label{glaw}
              p = (\gamma - 1)\rho \quad ,
\end{equation}
with the EFE, it is easily seen that the evolution equation for the 
scale function can be cast in the 
form below\cite{LGA}

\begin{equation}
\label{evolR}
     R\ddot{R} + (\frac{3 \gamma - 2}{2} - \frac{\gamma
     \psi}{2nH}) \dot{R}^2 + ( \frac{3 \gamma - 2}{2} -
     \frac{\gamma \psi}{2nH}) k = 0 \quad.  
\end{equation}

Consider now the following matter creation rate:

\begin{equation}
\label{defpsi}
\psi = 3 \beta n H  + 3({1-\beta})n\frac{{H}^{2}}{H_I}  \quad .
\end{equation}
The first term in the right-hand-side(RHS) of 
above equation is the same
creation rate 
considered in Ref.\cite{LGA}(from now on referred to as paper I). The 
second one is a correction term of the 
order of $\frac{H}{H_{I}}$. For $\beta=0$, Eq.(8) reduces to the 
matter creation rate considered by Lima and Germano\cite{LG}, the 
consequences of which have been analysed 
only in the flat case. The main difficulty of 
such a scenario is closely related with 
the predicted small time interval elapsed during   
the FRW phase. As shown in \cite{LGA},  
the presence of the $\beta$ parameter is crucial to 
solve the ``age problem'' 
since it parametrizes the extent to 
which the model departs 
from FRW phase at late stages.
The arbitrary time scale $H_{I}^{-1}$ 
characterizes the initial de Sitter 
phase and, together with the $\beta$ 
parameter, will presumably be given by some 
fundamental model of matter 
creation. At late times e.g., for $H << H_{I}$, the 
first term on the RHS 
of (8) is dominant. Hence, the model discussed here may also be 
viewed as an early phase of the ``adiabatic'' matter 
creation scenario 
proposed in the paper I. 
Using the above expression we can recast  Eq.(7) in the FRW-type form:

\begin{equation}
\label{evolRb}
     R\ddot{R} + (\frac{3 \gamma_* - 2}{2}) \dot{R}^2 + 
	(\frac{3 \gamma_* - 2}{2}) k = 0 \quad ,  
\end{equation}
where $\gamma_{*}$ is an effective time-dependent ``adiabatic index'' 
given by

\begin{equation}
\label{gind}
\gamma_* = \gamma (1-\beta) (1-\frac{H}{H_I}) \quad .
\end{equation}
For either $H=H_I$ or $\beta =1$, equation (10) gives $\gamma_{*}=0$, and 
(8) reduces to 

\begin{equation}
R{\ddot R}-{\dot R}^2 -k =0\ ,
\end{equation}
which yields the well known de Sitter solutions

\begin{equation}
R(t)=\left\{
\begin{array}{ll}
             H_I^{-1}\cosh(H_I t) & \ \ \ \ \mbox{$k=+1$}, \\
             R_* e^{H_I t} & \ \ \ \ \mbox{$k=0$} \ \ ,\\
             H_I^{-1}\sinh(H_I t) & \ \ \ \ \mbox{$k=-1$}. 
\end{array}\right.
\end{equation}
Hence, unlike in the standard FRW model, the present scenario
begins in a pure nonsingular de Sitter vacuum with Hubble 
parameter $H=H_I$ regardless the 
value of the curvature parameter. Accordingly, equation(2) 
gives $\rho=\frac {3{H_{I}}^{2}}{8 \pi {G}}$ as it 
should be due to the symmetries of 
the de Sitter spacetime.   
Analytically, the ansatz (8) can be viewed as the 
simplest matter creation law which destabilizes the initial de Sitter configurations given by (12).
In this way, the initial evolution is such 
that the singularity, flatness and horizon 
problems may be simultaneously
eliminated. All these asymptotic solutions have constant curvature,
and are unstable in the future. Of course, closed ($k=1$) solutions are not of the ``bouncing'' type, rather the universe begins its evolution from a closed de Sitter universe.

In the opposite limit, $H<< H_{I}$, Eq.(10) reduces to $ \gamma_{*} = \gamma (1-\beta)$ with (9) taking the following form

\begin{equation}
       R \ddot{R} + \Delta  \dot{R}^2 + \Delta k = 0     \quad,
\end{equation}

\noindent the first integral of which is

\begin{equation} 
{\dot{R}}^2 =  {\frac {B}{{R}
^{2\Delta}}} - k 
\quad ,
\end{equation}
\noindent where $\Delta = \frac{3\gamma(1-\beta)-2}{2}$ and 
$B$ is a positive constant (see eq.(2)). 

Using (14) one may express the energy density as well as the 
pressures ($p$ and 
$p_c$) as functions solely of the scale
factor $R$ and of the $\beta$ parameter. In fact, inserting (14) into 
(2), one obtains

\begin{equation}
\rho = \rho_o { ( \frac {R_o}{R} ) }^{3\gamma(1-\beta)} 
\quad , \end{equation}
where $\rho_o = 3B/8 \pi GR_{o}^{3\gamma(1-\beta)}$. 
The above equation shows that the densities of 
radiation and dust scale, respectively, as 
$\rho_{r} \sim R^{-4(1 - \beta)}$ and 
$\rho_{d} \sim R^{-3(1 - \beta)}$. Hence, the 
transition from a radiation to a dust dominated 
phase, in the course of the expansion, 
happens exactly as in the standard model. Of course, a more simplified scenario may also be implemented. For instance, if one assumes that 
only photons are produced
such a transition is even more smooth than in the 
FRW case, since the $\beta$ factor for the 
matter energy density($\gamma=1$) in (15) must be suppressed. Another possibility is to consider 
that the $\beta$ parameter is always nonnull, however, assuming 
different values when the universe evolves from 
radiation to a matter dominated phase. 

Summarizing, the cosmic history in the deflationary scenario
proposed here
proceeds in tree stages. Firstly, there is a natural transition
from an early de Sitter phase to 
a slightly modified FRW stage 
irrespective of the values 
of k and $\gamma$.
For $\gamma=\frac{4}{3}$, the model enters  smoothly in the radiation 
dominated phase providing an interesting solution to 
the ``exit problem'' of the old inflationary 
scenario. Subsequently, the transition to the present dust phase 
is completed in the same fashion of the standard 
model. Indeed, the quasi-FRW stage, which is defined by  
condition $H<<H_{I}$ correspond, at the level of the
matter creation rate (8),
to considering only the first term on the RHS of 
Eq.(8). This case  
has been studied in detail in paper I.
As shown there, the $\beta$ parameter
plays an important role during that phase 
since it increase
the age of these models. Note also that all curvature 
invariants are 
bounded for any instant of time, and as a consequence 
the geodesics of comoving
observers are unbounded. It thus follows that such spacetimes 
evolve from a nonsingular initial state. 
   
For simplicity and also to obtain an exact description we now 
consider $k=0$ as preferred by inflation. In terms 
of the Hubble parameter we can rewrite (9) as

\begin{equation}
\label{varH}
\dot H + \frac{ 3 \gamma (1-\beta)}{2} H^2 (1-\frac{H}{H_I}) = 0 \quad ,
\end{equation}
which is just the equation of motion for the flat vacuum decaying model of Ref.\cite{LimaM}. The 
solution of (16) can be written as 

\begin{equation}
\label{solH}
H = \frac{H_I}{1+ C R^{\frac{3\gamma(1-\beta)}{2}}} \quad ,
\end{equation}
where $C$ is a $\gamma$-dependent integration constant. Note that
$H=H_I$ is a special solution of (16) describing the de 
Sitter spacetime 
for any value of $\gamma$.
Such a solution is clearly unstable with respect to the
critical value $C=0$. For $C>0$, the universe starts without 
singularity 
and evolves continuously towards a quasi-FRW stage, $R \sim {t}^{\frac{2}{3\gamma(1-\beta)}}$, for large values of the 
cosmological time. It is also very easy to 
show that such a universe is
horizon free. in fact, a light pulse beginning at $t=-\infty$ will have traveled by the time t a physical 
distance $l_{H}(t)= R(t)\int_{-\infty}^{t}\frac{dt`}{R(t`)}$. From (17) we may
write

\begin{equation}
l_{H}(t)=H_{I}^{-1}R(t)
\int_{0}^{R(t)}\frac{(1 + CR^{3\gamma(1-\beta)/2})dR}{R^{2}}
\quad.
\end{equation}
Since the above integral diverge at the lower limit, the model is free of horizons e.g., light signals could have traveled to infinite distance 
at any t, since
the universe came into existence at coordinate time $t=-\infty$ thereby,
allowing the interactions to homogenize the whole universe. 

The dynamic qualitative behavior sketched above may be
described exactly. To show this, it proves convenient to 
compute C in 
terms of the 
present day parameters $H_o$, $R_o$ and also of $H_I$. From 
(17) one reads off

$$
C= (\frac{H_I - H_o}{H_o}) \frac{1}{R_o^{3\gamma (1-\beta)/2}} \quad ,
$$
and by recasting this constant in terms of

\begin{equation}
R_{*}=R_o (\frac{H_o}{H_I - H_o})^{2/3\gamma(1-\beta)} \quad ,
\end{equation}
which is determined by taking $H(R_*)=H_I/2$, one can write the 
integral of (17)
in the convenient form

\begin{equation}
\label{sol_R}
 H_I t=\ln(\frac{R}{R_*})  + 
 \frac{2(H_I - H_o)}{3\gamma(1-\beta)H_o} {(R/R_{0})}^{3\gamma(1-\beta)/2} 
\quad . 
\end{equation}
Once $H_I$ is given, scale $R_*(\gamma , \beta)$ automatically defines, for 
each model, 
the end of the deflation and 
the beginning of the quasi-FRW phase, as seen in Fig. 1 . At early times 
($R \ll R_{*}$) , when the logarithmic term is dominant, we obtain  

$$
 R \simeq  R_*e^{H_I t} \quad , 
$$
in accordance with Eq.(12). At late times ($R \gg R_*, H \ll H_I$) 
the expression reduces to the second term in (20)

\begin{equation}
R \simeq R_0 \left[3\gamma(1-\beta)\frac{H_0 
t}{2}\right]^{2/3\gamma(1-\beta)}\  \quad . \end{equation}
The above expression shows us how the $\beta$ parameter may 
be useful in reconciling 
the theoretically favored $\Omega_o=1$ with the latest measurements
of the Hubble parameter\cite{Pierce,Freedman}. In a matter 
dominated universe ($\gamma=1$),
the time interval ${\Delta}t=t_{o}$
elapsed in the quasi-FRW 
stage is $t_{o}\sim \frac {2{H_o}
^{-1}}{3(1 -\beta)}$, when in the standard
flat model $(\beta=0)$, one would obtain exactly
$\frac{2}{3}{{H_o}^{-1}}$ (see also Fig. 1). 
As remarked earlier, the arbitrary time scale ${H_I}^{-1}$ 
furnishes the greatest value of energy density 
$(\rho=\frac{3H_{I}^{2}}{8{\pi}G})$. In addition, it is readily seen that the
maximum  matter creation rate, $\psi_{I}=3n_{I}H_{I}$, also occurs in 
the de Sitter phase. In fact, by inserting 
(8) into (4), the balance equation for the particle number 
density assumes the form
 
\begin{equation}
\label{varn}
{\dot{n}} + 3n{(1- \beta)}
H({1- \frac{H}{H_I}})=0 \quad ,
\end{equation}
and from (17), the solution of the above equation may be cast as
\begin{equation}
n=n_{I}[1 + (\frac{R}{R_*})^{3\gamma(1- \beta)/2}]
^{-\frac{2}{\gamma}}   \quad .
\end{equation}
In the 
de Sitter phase, $R<<R_*$, the 
above equation 
yields a constant particle number density, $n \sim n_{I}$, and 
as a consequence the net number of particles N 
grows proportional to
$R^{3}$. In the opposite regime, $R>>R_*$ ($H<<H_I$), 
n scales with ${R}^{-3(1-\beta)}$ as expected
for the quasi-FRW stage (see paper I). In what follows, the above
results will be extensively used.

\section{Thermodynamic Behavior}

\hspace{.3in} Let us now discuss some 
thermodynamic features of the 
deflationary scenario proposed in the 
later section. For adiabatic matter creation 
the temperature law and the rate of variation 
of the specific entropy are given by\cite{CLW,LG}   

\begin{equation}
\frac{\dot T}{T} = {( \frac{\partial p}{\partial \rho} )} 
\frac{\dot n}{n} \quad ,
\end{equation}

\begin{equation}
\dot{\sigma} = 0 \quad ,
\end{equation}
where T is the temperature and $\sigma$ is the 
specific entropy (per particle).
Using the $\gamma-$law equation of 
state, a straightforward integration
of (24) yields

\begin{equation}
T = T_I {( \frac{n}{n_I} ) }
^{\gamma-1} \quad ,
\end{equation}
or still, inserting (23) 

\begin{equation}
T= T_I [ 1  +  (\frac{R}{R_*})^{3\gamma(1-\beta)/2}]^
{-2\frac{(\gamma-1)}{\gamma}} \quad ,
\end{equation}
where $T_{I}$ is the (maximal) temperature 
at de Sitter phase. Due to irreversible matter 
creation, the expansion proceeds isothermically during the de 
Sitter phase ($R<<R_*$), thereby 
avoiding the extremely 
rapid supercooling and the 
subsequent dramatic reheating that must take place 
at $R\simeq R_*$ in all inflationary 
scenarios\cite{TraschenRB}. The temperature 
decreases continuously in the course of 
the expansion. At late stages, 
when the universe
enters in the quasi-FRW phase, the temperature 
scales as $T \sim R^{-3(1- \beta)}$, as 
expected(see paper I). For $\beta=0$,
the result of Ref.\cite{LG} is recovered, with 
the universe evolving 
towards the standard FRW model. 
Now, recalling
that $R_*$ was determined in terms 
of $H_I$, $H_o$, $R_o$ and $\beta$ by (19), the 
above temperature evolution
law depends only on three unknown 
quantities, namely: Two initial conditions $T_I$ and $H_I$, plus 
the $\beta$
parameter. Note, however, that the $\beta$-dependence in (27) 
cancels at $R=R_o$ thereby, reducing our problem
to the right choice of the 
initial conditions. Since the model starts as a de 
Sitter spacetime, the most natural choice is to define $T_I$ as 
the Hawking temperature

\begin{equation} 
T_I=\frac{\hbar H_I}{2\pi k_B}\ ,
\end{equation} 
Of course, such a choice is dictated only by 
the overall existence of 
an initial de Sitter phase so that it could be 
applied irrespective
of the value of the curvature parameter. The  
arbitrary time scale of the de Sitter state, $H_I^{-1}$,
has not been  fixed by the model.
For consistency, since we are working 
with a classical description, 
 $H_I^{-1}$ must be chosen in such 
way that (28) becomes of the
same order or smaller than 
Planck temperature. Indeed, in the 
framework of quantum cosmology,
many authors have suggested that the 
spontaneous birth of the universe
leads naturally to a de Sitter 
stage with $H^{-1}\sim t_{pl}$ or
equivalently an energy 
density $\rho_ \sim \rho_{pl}$ 
(see for example \cite{AV 82}).
It is remarkable that such a choice, say, 
$H_I=\frac{2\pi t_{pl}^{-1}}{\alpha}$, where $\alpha$ is 
a pure number, allow us 
to estimate the present value of the temperature 
of CMBR. In fact, from (28), the initial temperature of 
the universe is close to Planck temperature

\begin{equation} 
T_I=\frac{1}
{{\alpha}k_B}\sqrt{\frac{\hbar}{G}} 
\sim \frac{1.4 \times 10^{32}}{\alpha} K   \quad . 
\end{equation} 
Now, taking $\gamma=4/3$ and
substituting $R_*$ into (27) we may write

\begin{equation}
T= T_I 
[1+{\frac{H_I - H_o}{H_o}} (R/R_o)
^{2(1-\beta)}]^{-\frac {1}{2}} \quad ,
\end{equation}
and since $H_I>>H_o$, replacing (29) and using
the previously defined value of $H_{I}$, 
we obtain at the present time

\begin{equation}
\label{T_o_r}
T_o = \frac {1.4 \times 10^{32}K}{{\sqrt \alpha}}
{(\frac{H_o t_{pl}}{2\pi})^{1/2}} \quad ,
\end{equation}
where $T_o$ is the temperature at $R=R_o$.
The above expression may be used either
to estimate $T_o$ if the 
value of $H_o$ is
given or to compute $H_o$ using the well 
established measurements of $T_o$.
For example, assuming that $H_o$ is centered 
at $80 \, km \, s^{-1} \, Mpc^{-1}$ 
as claimed by Freedman et al.\cite{Freedman}, this
means that $H_{o} \approx 2.7 \times 10^{-18}s^{-1}$  
and since $t_{P} \sim 5.4 \times 10^{-44}s$, we 
obtain $T_{o} \sim \frac{21 K}{\sqrt \alpha}$. 
Therefore, 
taking $\alpha$ of the order
of 50, it follows that $T_o \sim 2.8 K$ and 
from (28), the initial 
temperature of the 
universe is 50 times smaller 
than Planck temperature.

Let us now discuss the entropy 
production. By definition, the specific 
entropy(per particle) 
is $\sigma=\frac {S}{N}$, it thus follows  
from (24) that

\begin{equation}
\label{Sdot}
	\frac{\dot S}{S} = \frac{\dot N}{N} \quad ,
\end{equation}  
and since $\dot N \ge 0$ in all stages, the total 
entropy increases in 
the course of the evolution. Unlike of all 
microscopic variants of 
inflation, entropy 
generation in our scenario is concomitant 
with the overall deflationary process. To 
compute the net 
entropy in terms of the scale function,
it proves convenient to consider the 
entropy density $s_I$ of the de Sitter 
phase as an initial condition. In 
addition, instead of integrating (32), it is
much simpler to consider (23) together with the
constant value of $\sigma=\sigma_I$, to obtain  

\begin{equation}
S=\frac {s_{I}R^{3}}{\left[ 1 + (\frac{R}{R_*})^{3\gamma(1- \beta)/2} 
\right]^ {\frac{2}{\gamma}}}  \quad.
\end{equation}
Hence, during the de Sitter phase ($R<<R_*$) entropy 
scales as $S \sim R^{3}$, whereas 
at late stages $S \sim R^{3\beta}$ as it should (see paper I). In Fig. 2 
we have plotted energy density $\rho$, total entropy $S$ and temperature 
$T$ in the early stages of a radiation-dominated model with $\beta=0.25$.

For $\beta=0$, (33) reduces to equation (27) 
of Ref.\cite{LG}. In this case,
the entropy is constant at late times as 
expected for an adiabatic 
FRW type expansion. Now, inserting  
the value of $R_*$ given by (19) with $H_I >> H_o$,  
taking $\gamma={4/3}$ and ${R_o} \sim  H_{o}^{-1}$, to a 
high degree of accuracy (33) yields 
$S_o \sim s_I (\frac{H_I}{H_o})^{\frac{3}{2}}$. Of course,
the value of $s_I$ (or $n_I$) is uniquely defined in terms 
of $T_I$. Indeed, since the equilibrium relations are maintained, 
$\rho \sim {T}^4$ and 
$n \sim  T^{3}$ (see (26) and paper I), it 
follows that $s_I=\frac{4}{3}aT_{I}^{3}$, where 
$a=\frac {\pi^{2}}{15}$ is 
the radiation equilibrium 
constant ($\hbar=k_{B}=1$).
Finally, by replacing the value of $H_I$,  
the present radiation 
entropy in our horizon may be expressed as

\begin{equation}
\label{S_now}
S_o \approx  \frac {3.10^{-2}}{({\alpha H_o t_P})^{\frac{3}{2}}} \quad .
\end{equation}
Therefore, as we did in the case of temperature, taking $\alpha=50$ 
and  $H_o=80 \, km \, s^{-1} \, Mpc^{-1}$, we obtain for the present
dimensionless radiation entropy  $S_o \sim 2.9 \times 10^{87}$. 

\section{Conclusion}

\hspace{.3in}A class of deflationary cosmologies driven by ``adiabatic`` matter creation has been proposed. In these models, the beginning of the universe is a pure de Sitter state supported by
the matter creation process. Subsequently,
the universe evolves smoothly towards a radiation dominated 
phase as happens in the standard model. In this
way, both the singularity and horizon problems have 
been naturally solved. 
As a matter of fact, deflationary spacetimes seem to be a rather
generic solution of the EFE since it can also be generated 
by quite different
mechanisms like bulk viscosity\cite{Barrow} and vacuum decaying energy
density\cite{LimaM}. 

Unlike all variants of inflation, in such models
there is no supercooling or reheating. The temperature
of radiation diminishes slowly from its greatest value during the  
initial de Sitter configuration until the present $2.8 K$. 
Simultaneously, the entropy increases continuously to
reach $S_o \sim 10^{87}$ at the present phase. In addition, due to 
persistent matter creation the 
age of the universe also fits well with a higher 
value of $H_o$. More 
important, such 
results  were obtained using one and the same initial condition. In this 
connection, we 
remark that the initial Hawking-Gibbons temperature given by (28) is
not related here with the horizon temperature of a vacuum de Sitter 
spacetime.
Rather, this de Sitter phase is supported by the negative 
creation pressure and 
for $\gamma=\frac{4}{3}$ space is filled with radiation.
At late times, the model evolves to the scenario 
proposed in Ref.\cite{LGA}, thereby solving the ``age 
problem''. As argued there(see also
\cite{JA 96,JA1 96}), such a scenario is also 
compatible with the constraints from the cosmic background radiation, 
and a crucial test for these models is provided by 
the measurements of the temperature of the universe at high redshifts.

\section{Acknowledgments}
\vspace{5mm} 
It is a pleasure to thank Robert Brandenberger and Richhild Moessner for a critical reading of the manuscript. Many 
thanks are also due to Andrew
Sornborger and M. Trodden for their permanent stimulus and interest in this work. One 
of us (JASL) is grateful for the hospitality of the Physics Department
at Brown University. This work was partially supported by the Conselho 
Nacional de Desenvolvimento Cient\'{\i}fico e Tecnol\'{o}gico - CNPq 
(Brazilian
Research Agency), and by the US Department of Energy under grant 
DE-F602-91ER40688, Task A.

\pagebreak

{\large Figure Captions}

Figure 1:
\begin{quote}
{\footnotesize {\bf Fig. 1} - Scale factor as a function of time for 
$\gamma = 4/3$ (radiation) and $\beta= 0$ (left curve), 
$\beta=0.25$,
$\beta=0.5$ and $\beta=0.75$ (right curve). For $\beta > 0$, the universe deflates smoothly from an initial de Sitter phase
($R \ll R_{*}$) to a quasi-FRW stage($R \gg R_{*}$). 
The dashed curve 
depicts scale factor of
a FRW universe starting at $t=0$. 
When $\beta=0$ we have no matter creation at the
present time, and for $\beta \rightarrow 1$ the model inflates
at all times powered by a steady creation rate. The 
physically meaningful ages  
are easily seen to be the time elapsed from $t=0$ to 
the moment when a curve crosses a fixed scale($R=R_o$). As expected, models with larger
matter creation at late times (larger $\beta$) are older. }
\end{quote}

Figure 2:
\begin{quote}
{\footnotesize {\bf Fig. 2} - Energy density, total entropy and 
temperature of radiation as a function of the scale factor (normalized to 
$R_*$) in a scenario with $\beta=0.25$. 
Notice the nearly constant energy density and temperature, but a 
huge growing entropy during the deflationary process. As soon as
the scale $R=R_*$ is reached, regimes change and the 
FRW-like 
epoch, characterized by falling energy density and temperature, starts. 
The model does not evolve adiabatically. The total entropy always increases, although moderately nowadays. } 
\end{quote}

\end{document}